\documentclass[sigconf]{acmart}
\usepackage{graphicx} 
\usepackage{todonotes}
\usepackage{enumitem}
\usepackage{amsmath}
\usepackage{soul}
\usepackage{circledsteps}

\setcopyright{acmcopyright}
\copyrightyear{2024}
\acmYear{2024}
\acmDOI{XXXXXXX.XXXXXXX}

\acmConference[MSR'24]{Mining Software Repositories}{2024}{Lisbon}
\acmPrice{15.00}
\acmISBN{978-1-4503-XXXX-X/18/06}

\begin{document}

\title{Comparison of Static Analysis Architecture Recovery Tools\\for Microservice Applications}

\author{Simon Schneider}
\affiliation{%
  \institution{Hamburg University of Technology}
  \country{Hamburg, Germany}
}

\author{Alexander Bakhtin}
\affiliation{%
  \institution{University of Oulu}
  \country{Oulu, Finland}
}

\author{Xiaozhou Li}
\affiliation{%
  \institution{University of Oulu}
  \country{Oulu, Finland}
}

\author{Jacopo Soldani}
\affiliation{%
  \institution{University of Pisa}
  \country{Pisa, Italy}
}

\author{Antonio Brogi}
\affiliation{%
  \institution{University of Pisa}
  \country{Pisa, Italy}
}

\author{Tomas Cerny}
\affiliation{%
  \institution{University of Arizona}
  \country{Tucson, USA}
}

\author{Riccardo Scandariato}
\affiliation{%
  \institution{Hamburg University of Technology}
  \country{Hamburg, Germany}
}

\author{Davide Taibi}
\affiliation{%
  \institution{University of Oulu}
  \country{Oulu, Finland}
}
\affiliation{%
  \institution{Tampere University}
  \country{Tampere, Finland}
}

\begin{abstract}
Architecture recovery tools help software engineers obtain an overview of their software systems during all phases of the software development lifecycle.
This is especially important for microservice applications because their distributed nature makes it more challenging to oversee the architecture.
Various tools and techniques for this task are presented in academic and grey literature sources.
Practitioners and researchers can benefit from a comprehensive overview of these tools and their abilities.
However, no such overview exists that is based on executing the identified tools and assessing their outputs regarding effectiveness.
With the study described in this paper, we plan to first identify static analysis architecture recovery tools for microservice applications via a multi-vocal literature review, and then execute them on a common dataset and compare the measured effectiveness in architecture recovery.
We will focus on static approaches because they are also suitable for integration into fast-paced CI/CD pipelines.

\end{abstract}

\keywords{microservices, static analysis, architecture recovery}

\maketitle

\section{Introduction}
Static analysis tools can support developers with valuable feedback on their work without the need to run and test their systems. 
Tools such as SonarQube~\footnote{sonarsource.com/products/sonarqube/}, PMD~\footnote{pmd.github.io/}, or IntelliJ~\footnote{jetbrains.com/de-de/idea/} are examples of widely popular solutions that perform real-time analyses for different aspects of development.
Static analysis architecture recovery tools can support developers by showing the implemented system's high-level architectural design, helping them adhere to the intended design and avoid architectural issues such as violations of security rules~\cite{Tukaram22_security_rules}, bad smells~\cite{Ponce2022_MLRSecuritySmellsMSA}, antipatterns~\cite{Taibi20_microservices_antipatterns}, or non-conformances~\cite{Cao24_catma}.
Providing accessibility of the systems' architecture is additionally important for distributed systems, where it is challenging for developers to oversee the complete application. 

The microservice architecture is an architectural style that often entails a distributed codebase.
It has been increasingly adopted in the last years and continues to gain popularity in software development.
Applications employing the microservice architecture split their business logic into multiple microservices. 
The individual microservices communicate over lightweight communication channels~\cite{Dragoni16_microservices_yesterday_today_tomorrow, Lewis14_microservices}.
The architecture has many benefits for software engineering activities.
However, the distributed nature of the codebase also poses challenges, since it is more difficult to gain and maintain an overview of the application's architectural design~\cite{DiFrancesco19_architecting_microservices, Soldani2018_MSAPainsGains}.

Architecture recovery tools and techniques can support developers and analysts in this regard by creating a representation of the implemented system.
In the field of program comprehension, it has been shown that such representations foster better and easier analysis, maintainability, and usability, and support software engineers during development (e.g.,~\cite{Arisholm06_impact_uml, Budgen11_uml_slr, Gravino10_empirical_investigation_code_comprehension, Gravino15_code_comprehension_uml, Schneider24_DFDs_empirical_experiment}).
By allowing to assess the adherence of the implemented architecture to the designed one, issues such as architectural drift are also mitigated.

With the growing adoption of the microservice architecture, the need for static analysis tools that specialize in microservice applications rises.
Consequently, various approaches for architecture recovery for microservices have been proposed in academic literature~\cite{Alshuqayran18, Granchelli17, Kleehaus18, Queval23_extracting_architecture, Soldani21}.
Some authors also provide tools to show the feasibility of their presented approaches.
Further tools can be found in the grey literature.

In this paper, we present a study that we plan to conduct to identify and compare static analysis tools for architecture recovery of microservice applications.
A core contribution is the execution of all identified tools on a common dataset and the comparison of their effectiveness in performing architecture recovery.
We measure effectiveness in terms of precision and recall of extracted application characteristics, compared to a manually created ground truth.

In the context of this work, we refer to a microservice application's architecture as a representation of its architectural design in terms of components and their logical connections.
The microservice architecture offers such a system decomposition by definition since the individual microservices are meant to be self-standing, independently deployable units.
The communication links between them that are necessary to fulfill the system's business logic form the connections between components.

The results we expect to obtain from our study can be beneficial to both researchers and practitioners. 
For researchers, an overview of existing tools can prevent the creation of yet another approach for which a similar technique has already been proposed.
Consequently, the results of the study can shed light on the directions for future work and help accelerate research on the topic.
For practitioners, the results of the planned study can provide a reference for the tools and for comparing their actual capabilities.
Industry adoption of tools and techniques presented in academic literature is notoriously challenging.
Our study is geared towards fostering visibility of tools for microservice architecture recovery and showing their effectiveness in basic architecture recovery (i.e., the extraction of components and connections) as well as additional characteristics.

\section{Related Work}

Several literature reviews have been presented that address domains similar to those of our planned study.
For example, Dragoni et al.~\cite{Dragoni16_microservices_yesterday_today_tomorrow} and Soldani et al.~\cite{Soldani2018_MSAPainsGains} conducted literature reviews for the microservice architecture in general.
Others present reviews of approaches for a wide variety of specific use cases.
For example, Abdelfattah et al.~\cite{Abdelfattah_2023} present a review of approaches for reconstruction, reasoning, and evolution, and Bushong et al.~\cite{Bushong_2021} a review of reconstruction, architectural degradation, and technical debt, among others.
Neri et al.~\cite{Neri2020_MLRPrinciplesSmellsRefactoringsMSA} and Ponce et al.~\cite{Ponce2022_MLRSecuritySmellsMSA} conducted literature reviews with more narrowly defined scopes, both focusing on smells that are specific to microservice applications.
Fritzsch et al.~\cite{Fritzsch19_refactoring_review} presented a review of microservice refactoring approaches for migration from monolithic to microservice architecture.
Gortney et al.~\cite{Gortney22_visualizing_microservices} presented a review of approaches for microservice reconstruction.
The authors focused on dynamic analysis approaches, which are based on, e.g., log analysis, tracing technologies, or monitoring technologies.
Finally, Cerny et al.~\cite{Cerny22_microservice_reconstruction_review} also presented a review of microservice architecture reconstruction approaches without restricting the search to dynamic approaches.
However, different from our planned study, these publications all present literature reviews of techniques instead of tools, meaning, that they do not consider whether the found approaches are supported by implementations. 

Some work has also been published on reviews of tools. 
Although not focussed on microservice applications, Emanuelsson and Nilsson~\cite{EMANUELSSON20085}, Lenarduzzi et al.~\cite{LENARDUZZI2023111575}, and Mantere and Uusitalo~\cite{Mantere09_comparison_static_analysis_tools} compared static analysis tools for different use cases.
Bakhtin et al.~\cite{Bakhtin22_api_patterns} performed a systematic grey literature review of tools detecting microservice API patterns, and Giamatti et al.~\cite{GIAMATTEI2024111906} presented the results of a systematic grey literature review of monitoring and DevOps tools for microservice applications, thus focusing on the same domain we intend to target with our planned study.

The planned study is based on and extends the work presented by Bakhtin et al.~\cite{Bakhtin23_tools_study}. 
The authors conducted a systematic mapping study based on the academic literature to identify tools for the architecture recovery of microservice applications.
The review is based on information reported by the corresponding publications, and not on insights gained from executing the tools, which we plan to do instead.
Reviewing tools based on their reported evaluations is insufficient for a comprehensive comparison because they are not applied to the same dataset and further, they are often not evaluated thoroughly, as noted by Schneider and Scandariato~\cite{Schneider23_code2dfd} as well as by Akkaya and Ovatman~\cite{akkaya2022comparative}.
Instead, evaluations are often based on executing the corresponding tool on less than five applications.

Notably, Akkaya and Ovatman~\cite{akkaya2022comparative} performed a similar study to the one we propose.
They identified three tools for microservice extraction in the literature and executed them on four applications.
However, they only considered the detection of microservices and neglected all other characteristics.
Consequently, they do not review a holistic architecture recovery process, as we intend to do.

In conclusion, to the best of our knowledge, no review of static analysis microservice architecture recovery tools has been presented in the literature that provides a comprehensive comparison of the tools.
With the planned study, we aim at addressing this gap in the literature and provide a comprehensive comparison of such tools that can be found in academic as well as grey literature.
The comparison shall be based on the tools' observed effectiveness in architecture recovery on a shared dataset.

\section{Objectives and Research Questions}

In this work, we aim to provide researchers and practitioners with an overview of the state-of-the-art, freely available, static analysis architecture recovery tools for microservice applications.
Insights on the quality of such tools, as well as on the specific characteristics they extract, are valuable for both groups of stakeholders.
With the comparison of the tools' effectiveness in architecture recovery instead of an overview of their approaches, our study fills a gap in the literature that has not been addressed before.
Especially in academia, where published tools are often prototypes created for the sake of showing the feasibility of a presented approach and where subsequent maintenance is often neglected, such an evaluation is crucial for properly judging the tools' qualities.

In pursuing to fulfill the above objectives, we will answer the following research questions:

\begin{itemize}[leftmargin=*]
    \item \textbf{RQ1: Which freely available, static analysis tools for architecture recovery of microservice applications exist?} \\
    Several architecture recovery approaches have been proposed in the literature, which are often supported by prototypes implementing the techniques.
    Additionally, grey literature sources give pointers to further tools that fit this scope.
    We will curate a list of such tools in a multivocal literature review, which will then be evaluated and compared on a common benchmark.
    
    \item \textbf{RQ2: Which characteristics do the tools extract in addition to the basic architecture (i.e., components and connections between them)?} \\
    All architecture recovery tools extract the analyzed application's basic architecture (which, in the context of our work, are the application \emph{characteristics} services and connections between them).
    Many tools extract additional application characteristics, i.e., properties beyond the basic architecture, such as information about implemented security mechanisms, links to design requirements, or trust boundaries.
    An overview of these additional characteristics helps to identify tools for specific use cases.
    
    \item \textbf{RQ3: Which are the most commonly considered characteristics extracted by the tools?} \\
    Based on the presentation of the characteristics extracted by the identified tools, we will analyze the tools' overlaps and differences in their extraction scopes.
    The results will show what the tools mostly focus on and also where possible gaps lie.
    
    \item \textbf{RQ4: Which is the identified tools' effectiveness in architecture recovery?} \\
    To compare the identified tools based on their effectiveness in architecture recovery, we will execute them on a common benchmark and measure their precision and recall concerning the following properties:
        \begin{itemize}[leftmargin=*]
            \item \textbf{RQ4.1: Which is the identified tools' effectiveness in detecting the components that form a microservice application?} \\
            The individual microservices of an application constitute the building blocks that the application's microservice architecture consists of.
            We follow the distinction of Schneider et al.~\cite{Schneider23_code2dfd}, which says that internal microservices realize the application's business logic and are implemented specifically for an application.
            Infrastructural microservices instead use mainly existing code libraries that are  adjusted to the application's requirements (e.g., API gateways, authorization servers, or message brokers).
            The components are often the easiest characteristics to extract for microservice applications since deployment technologies such as Docker Compose or Kubernetes ease their detection.
            Nevertheless, this is the foundational step of architecture recovery and will thus be evaluated.
            
            \item \textbf{RQ4.2: Which is the identified tools' effectiveness in detecting connections between the components forming a microservice application?} \\
            The second group of application characteristics in the core architecture of microservice applications is that of logical connections between the components over which requests are made and data is exchanged.
            Such connections can be realized in different ways, for example via direct API calls, asynchronous communication techniques, or implicit invocations by infrastructural components such as the communication for registering services in a service registry.
            Due to this added complexity, the detection of connections is harder than that of the components and will be evaluated separately.
            \item \textbf{RQ4.3: Which is the tools' effectiveness in detecting the additionally extracted characteristics?} \\
            For those tools that extract characteristics in addition to the basic architecture (see RQ2), we will measure their effectiveness in extracting this extra information.
            We will directly compare tools that extract the same additional characteristics.
        \end{itemize}
\end{itemize}

\section{Methodology}
\label{sec:methodology}

\begin{figure*}
    \centering
    \includegraphics[width = 0.9\linewidth]{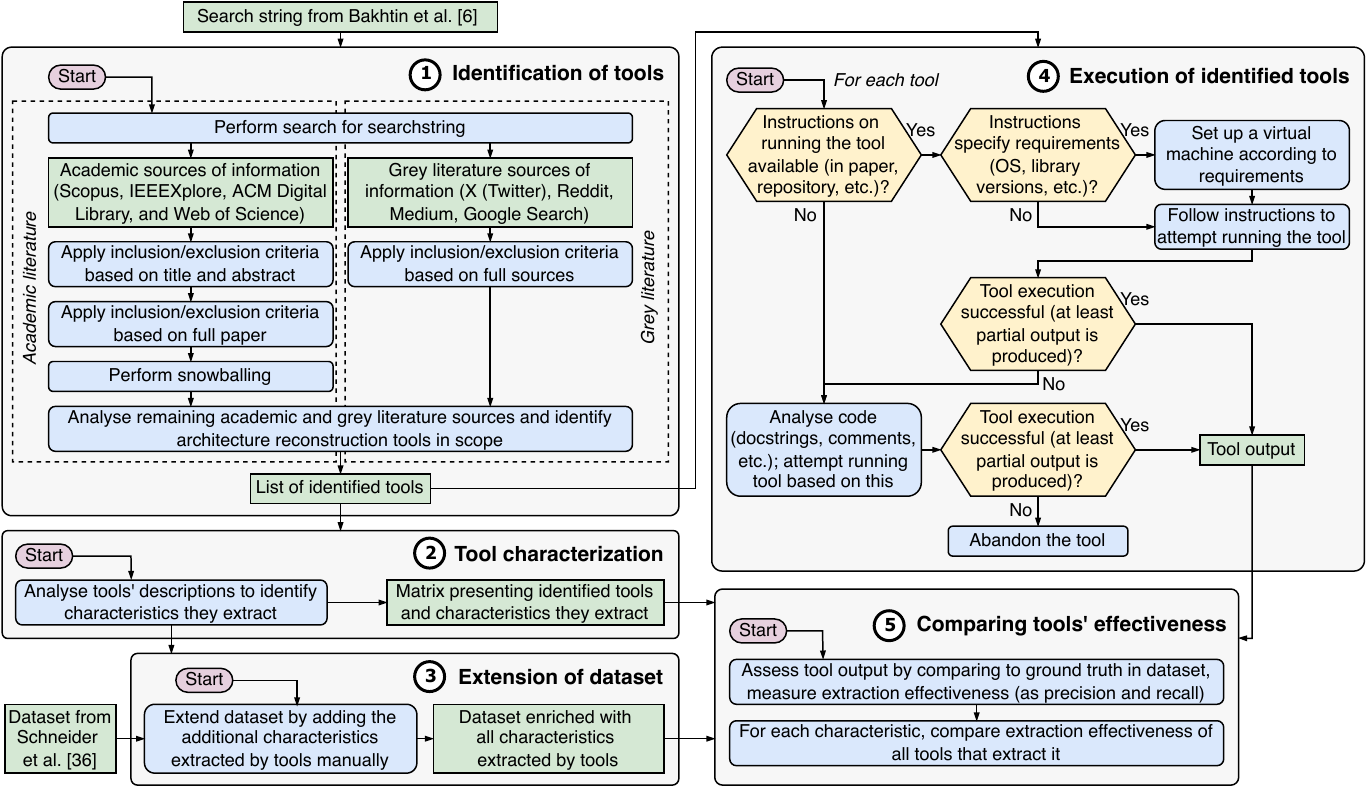}
    \vspace{-3mm}
    \caption{Methodology of the planned study.}
    \vspace{-4mm}
    \label{fig:methodology}
\end{figure*}

The planned study consists of broadly two stages, (i) performing a multi-vocal literature review to identify static analysis architecture recovery tools for microservice applications, and (ii) comparing the identified tools' effectiveness in architecture recovery by executing them on a common dataset and evaluating the outputs they produce.
Figure~\ref{fig:methodology} shows the complete methodology of the planned study structured into five steps.
Each step is further described in the following sections, in the order indicated by Figure~\ref{fig:methodology}.

\vspace{-2mm}

\subsection{Identification of Tools} 
\label{sub:methodology_tool_selection}

We will repeat the literature review of academic sources performed by Bakhtin et al.~\cite{Bakhtin23_tools_study} to identify tools published after the authors performed their search.
Further, we will apply our in- and exclusion criteria to the tools already identified by the authors to select those relevant to us.
Since the scope of tools considered in the planned study is a subset of the ones identified by Bakhtin et al.~\cite{Bakhtin23_tools_study}, and since the methodology we apply is adapted from theirs, we can assume that no relevant tools will be missed in this process.
Additionally to academic sources, we will adopt the methodology and apply it to grey literature sources as well, thus extending the work into a multivocal literature review~\cite{GAROUSI2019101} (see step~\Circled{1} in Figure~\ref{fig:methodology}).

For the repetition of the literature review of Bahtin et al.\cite{Bakhtin23_tools_study}, we will use the same search string (given below) and search for it in the same four scientific databases (\textsc{Scopus},\footnote{The \textsc{Scopus} database: \url{https://www.scopus.com}.}, \textsc{IEEEXplore}\footnote{The \textsc{IEEEXplore} database: \url{https://ieeexplore.ieee.org/}.},  \textsc{ACM Digital Library},\footnote{The \textsc{ACM Digital Library}: \url{https://dl.acm.org}.} and \textsc{Web of Science}\footnote{\textsc{Web of Science}:  \url{https://www.webofscience.com/wos/woscc/basic-search}}).
We will only consider results published after their reported search date.

\begin{center}
\small
\ttfamily
(Microservice* OR Micro-service* OR "micro-service*") \\
AND Architect* \\
AND (Reconstr* OR Mining OR Reverse engineering \\ OR Recover* OR Extract* OR Discover*) \\
AND (Tool* OR Prototype OR Implementation OR GitHub OR \\ Proof of concept OR POC OR Proof-of-concept)
\end{center}

\noindent
The fourth AND-group of the search string is used for in-text search if the database functionality allows it.

For grey literature sources, we will apply the search string to the four popular grey literature websites \textsc{Google Search}\footnote{\textsc{Google Search}: \url{https://google.com}}, \textsc{Twitter (X)}\footnote{\textsc{Twitter (X)}: \url{https://x.com}}, \textsc{Reddit}\footnote{\textsc{Reddit}: \url{https://reddit.com}}, and \textsc{Medium}\footnote{\textsc{Medium}: \url{https://medium.com}}.
These are popular websites and are used as sources of grey literature by other reviews, such as \cite{Moreschini2023TowardEM, PELTONEN2021106571}.
Duplicates will be removed during result aggregation.

After compiling the list of initial sources, we will apply the following inclusion and exclusion criteria, adapted from \cite{Bakhtin23_tools_study}:
\begin{itemize}[leftmargin=*]
    \item Inclusion criteria
    \begin{itemize}[leftmargin=*]
        \item Mentions a tool for microservice architecture recovery
        \item A reference to the freely available tool is made or the tool can be found by searching for its name
        \item The tool follows a static analysis or hybrid approach
    \end{itemize}
    \item Exclusion criteria
    \begin{itemize}[leftmargin=*]
        \item Source is not in English
        \item Out of topic - relevant terms are used in a different context
        \item Source describes different aspects of microservice recovery (not dealing with tools)
        \item Source describes closely related tasks such as monolith to microservice migration
    \end{itemize}
\end{itemize}

\noindent
For the academic sources, we will apply the criteria in two phases, as it is common practice when conducting SLRs.
In the first phase, sources are excluded based on reading the title and abstract of the paper; in the second phase, the complete paper is examined. 
For the grey literature sources, we will apply the criteria on the complete sources directly while also assessing the sources' quality regarding the producer's authority, the applied methodology, objectivity of the source, the date of publishing, and novelty, as per \cite{GAROUSI2019101}. If sources are deemed to be of insufficient quality, they will be excluded independent of their adherence to the inclusion criteria.

All steps will be performed by two authors independently, and disagreements solved via discussion with a third author.
How well authors agree with each other will also be analyzed with Cohen's kappa coefficient \cite{10.1023/A:1009820201126}. The kappa coefficient considers the observed frequency of agreement relative to the expected probability of agreement, assuming authors decide randomly and independently.

Finally, we will perform forward and backward snowballing~\cite{Wohlin} on the academic sources, i.e., reviewing all references and citations that occur in relevant places of the paper in the same way described above.
For the grey literature sources, we will instead check whether contained links to other resources refer to tools that are in scope.

We will examine the identified sources in detail to identify all presented and mentioned tools for microservice architecture recovery.
Specifically, we will look for any references to source code repositories, web applications, Docker images, or other ways of providing a tool.
In line with the objective of the study, only tools that follow a static analysis approach or hybrid approach where the static part can be run independently are considered.
The identification will be performed by the first author. 
In cases where no tools will be found in a source, another author will check the source as well for confirmation.
The resulting list of static analysis microservice architecture recovery tools will serve as the answer to \textbf{RQ1}.

\vspace{-2mm}

\subsection{Tool Characterization}
\label{subsub:tool_codification}

For all identified tools, we will further extract from the sources and from the resource where the tool is provided (e.g., the source code repository) their general properties (platform, language, static/hybrid approach, output format, etc.), as was done by Bakhtin et al.~\cite{Bakhtin23_tools_study} (step~\Circled{2} in Figure~\ref{fig:methodology}).
Here, those characteristics extracted by the tools that go beyond the basic architecture of the analyzed systems are of particular interest.
The results will be presented in a matrix listing all identified tools as well as the characteristics they extract.
This matrix will later determine which tools are compared with each other based on each characteristic. 
To this end, we expect to generalize some characteristics to allow a comparison between tools. 
For example, if one tool extracts information about implemented authorization mechanisms while others extract information about other security features, these characteristics will all be generalized into a common group of \emph{security mechanisms}.
In addition to guiding the comparison of tools' extraction effectiveness, the created matrix will also be the basis to answer \textbf{RQ2} and \textbf{RQ3}.

\vspace{-2mm}

\subsection{Extension of Dataset}
\label{sub:extension_of_dataset}
To compare the identified tools' correction in architecture recovery under controlled circumstances, they need to be executed on the same dataset.
We will use a dataset of 17 dataflow diagrams (DFDs) of open-source microservice applications~\cite{Schneider23_microsecend} for this purpose.
The DFDs depict the corresponding applications' architecture as well as additional properties. 
Their nodes represent the application's components, i.e., (internal and infrastructural) microservices, databases, and external entities; their edges represent connections between any two components.
As such, the nodes and edges are used as ground truth for the basic architecture (i.e., RQ4.1 and RQ4.2 will be answered based on the tools' effectiveness in extracting these characteristics).
The applications corresponding to the DFDs in the dataset are typical, small- to medium-sized open-source microservice applications written in Java with a focus on the Spring framework.
On average, the DFDs contain 11 nodes and 22 edges.
According to reports, Java is the most popular language for developing microservice applications~\cite{JetBrains22} and Spring is the most used framework for Java microservice applications~\cite{JRebel22}.
The applications in the dataset were selected from sources in the literature as well as popular repositories on GitHub. 
According to the authors, established design patterns for microservice applications using the Java Spring framework are prevalent in the applications.

Concerning the additional characteristics extracted by the tools (the basis for RQ4.3), the DFDs in the dataset contain extensive annotations that represent security mechanisms, deployment information, and other system properties (on average, 84 annotations per DFD).
We expect that this information will be in line with some of the tools' extracted characteristics but that it will not be sufficient to serve as ground truth for all identified tools.
Likely, some of them extract characteristics not currently contained in the DFDs.
For these cases, we will manually extend the dataset with the required information (step~\Circled{3} in Figure~\ref{fig:methodology}).
The methodology's details for how to detect the additional characteristics in the code depend on what exactly is to be extracted.
In general, at least four authors will take part in manually extracting the characteristics and cross-validating their results.
For this, the process will be performed by multiple authors independently.
Then, possible discrepancies between the authors' results will be solved by discussion with another author.
The extraction will be based on a manual analysis of the source code of the applications.
For this step, we will define code artifacts that indicate the existence of the characteristics for each of them.
These indicators will guide the manual extraction.
They will be formulated based on the tools' descriptions and the corresponding publications (for academic sources).

A further extension of the dataset will be needed in case tools are identified that analyze applications not written in Java. 
In this case, we will manually create DFDs for applications in the required language.
For this, we will follow the same process for creating the DFDs in the dataset described by Schneider et al.~\cite{Schneider23_microsecend}.
However, from our experience and the feasibility study (see Section~\ref{sub:feasibility_study}), most tools focus on Java.
This step might thus not be required.

As a result of the described process, an extended dataset will be an additional contribution of the planned work.
It will contain extensive information on different characteristics that are useful to benchmark a variety of different architecture recovery approaches.

\vspace{-2mm}

\subsection{Execution of Identified Tools}
\label{sec:execution}
To obtain outputs from the identified tools for their evaluation, we will run them on the applications in the extended dataset.
Naturally, the tools will need to be executed successfully to create outputs.
It is possible that there will be obstacles in terms of reproducibility, i.e., that it will not be trivial to execute some of the tools.
To achieve a fair comparison, a methodology for executing them was established (step~\Circled{4} in Figure~\ref{fig:methodology}) that ensures that the effort invested into attempting to run each tool will be comparable.
We will first attempt to run a tool based on available instructions (documentation, information in the source, etc.), possibly on a virtual machine if specific requirements for the execution environment are mentioned.
If this will not be successful, we will analyze the code for indicators of how to run the tool (code comments, hints by identifiers, etc.).
If all steps fail, we will abandon the tool and exclude it from the comparison.
As an indicator of successful execution, we will check whether the tool produces any output for any model item it is supposed to extract and which is present in the analyzed application.

\vspace{-2mm}

\subsection{Comparing Tools' Effectiveness}
We will use precision, recall, and execution time as quantitative measures for comparison.
These are common and objective metrics used for such evaluations.
Although we do not dictate a specific use case for the tools, their ability to perform their core functionality correctly is the most important basis for evaluation.
A tool should extract all existing characteristics it is supposed to detect and not falsely produce results for more than these.
Precision and recall serve as measures to indicate these two aspects.
The execution time is more dependent on the intended use case but is important for most scenarios as well.
Lightweight static analysis tools that show quick execution times lend themselves to being integrated into automated pipelines such as fast-paced CI/CD pipelines.
For example, the output of architecture recovery tools could be used by model-based analysis tools in a deployment pipeline.

To quantify the tools' output, we will manually count (step~\Circled{5} in Figure~\ref{fig:methodology}), by comparing to the ground truth, the number of correctly extracted characteristics (true positives, TP), the number of falsely extracted characteristics (false positives, FP), and the number of undetected characteristics (false negatives, FN).
Since the tools have different output formats, quantifying the results manually is deemed the safest method for a correct representation.
The process will be performed by two authors independently and disagreements solved in discussion with a third author.
Precision and recall will be calculated from these measures with the following formulas:

\vspace{-1mm}

\small
\[\text{Precision} = \frac{\text{Correct characteristics}}{\text{Correct characteristics}+\text{False characteristics}}\]
\vspace{-2mm}
\[\text{Recall} = \frac{\text{Correct characteristics}}{\text{Correct characteristics}+\text{Undetected characteristics}}\]\newline
\normalsize

\vspace{-5mm}
We will answer \textbf{RQ4} based on the above measures.
Specifically, we will compare different subsets of the complete list of tools against each other.
The overview of each tool's extracted characteristics (see Section~\ref{subsub:tool_codification}) will determine which tools are compared with each other. 
For each characteristic, we will compare the effectiveness achieved by all tools that are supposed to extract it in extracting this characteristic.
All tools will be compared on the characteristics \emph{components} and \emph{connections} since these form the basic architecture.

\vspace{-2mm}
\begin{table}[!ht]
    \footnotesize
    \centering
	\caption{Static analysis architecture recovery tools identified in the feasibility study. Pub. = Publication; Upd. = last update}
    \vspace{-3mm}
	\label{tbl:tools}
    \begin{tabular}{llcc}
        \toprule
        \textbf{Name} & \textbf{GitHub Repository} & \textbf{Pub.} & \textbf{Upd.}\\
        \midrule
        Attack Graph Generator & tum-i4/attack-graph-generator & \cite{Ibrahim19_attack-graph-generator} & 01/2021\\
        Code2DFD & tuhh-softsec/code2DFD & \cite{Schneider23_code2dfd} & 02/2024 \\
        MicroDepGraph & clowee/MicroDepGraph & \cite{Rahman19_microdepgraph} & 11/2021 \\
        microMiner & di-unipi-socc/microMiner & \cite{Muntoni21_microminer} & 11/2020 \\
        microTOM & di-unipi-socc/microTOM & \cite{Soldani23_microtom} & 01/2023 \\
        Prophet & cloudhubs/prophet & \cite{Bushong21_prophet} & 08/2023 \\
        RAD & cloudhubs/rad-analysis & \cite{Das21_rad} & 01/2021 \\
        \bottomrule
    \end{tabular}
    \vspace{-4mm}
\end{table}

\section{Feasibility Study}
\label{sub:feasibility_study}

To ensure the feasibility and insightfulness of the planned study, we conducted a small preliminary study.
First, we selected all static analysis microservice architecture recovery tools from the list of tools presented by Bakhtin et al.~\cite{Bakhtin23_tools_study}.
Table~\ref{tbl:tools} shows the seven tools that fit the inclusion criteria of the planned study.
Secondly, we attempted to run these tools following the presented methodology (see Section~\ref{sec:execution}).
We were successful in executing five of them within minutes.
The other two require a deeper analysis of the code (as specified in the methodology).
This step was omitted for now.
Without measuring the tools' effectiveness, this small preliminary study ensures the availability of data for comparison.
Note that the \emph{Attack Graph Generator} tool only performs architecture recovery as a means to achieve the generation of attack graphs. 
However, it produces an architectural representation as an intermediate result. 

\section{Potential Risks}
\label{sec:unknowns}

Some unforeseeable factors could influence the planned study.
They will have to be addressed when and if they manifest. 
We present here factors that we identified and how they could be addressed: 

\vspace{-1mm}
\paragraph{Inability to Execute Tools}
By design, the planned study relies on the successful execution of all identified tools.
It is possible that we will not be able to do so.
Some tools could show a lack of reproducibility that does not allow us to execute them successfully.
This is a realistic risk, especially for academic tools, which are often created as prototypes to prove the feasibility of presented approaches and not further maintained afterward.
As a consequence, we performed the feasibility study (see Section~\ref{sub:feasibility_study}) to verify that at least some tools are executable and will be part of the comparison.

\vspace{-1mm}
\paragraph{Tools' Extraction Scopes too Distinct for Comparison}
The overview of existing tools, the characteristics they extract, and their effectiveness in doing so is of high value on its own.
However, a core contribution of the planned study lies in comparing the tools' effectiveness.
If the tools are too distinct in their extraction scopes, we will present their effectiveness individually for the additional characteristics and will focus on comparing the results for nodes and edges, characteristics that all tools should extract by definition.

\vspace{-1mm}
\paragraph{Characteristics not existent in Dataset}
We may find tools that extract characteristics that are not contained in the initial dataset and that are too profound to be addressed by an extension of the dataset.
As described in Section~\ref{sub:extension_of_dataset}, we will extend the used dataset to serve as ground truth for all identified tools.
However, the additional characteristics may not occur in the applications in the dataset.
In such cases, we will assess the feasibility of extending the dataset with additional applications that show the missing characteristic.

\section{Threats to Validity}

Some threats to validity will apply to the planned study's results~\cite{Wohlin12_experimentation_in_se}.

\vspace{-1mm}
\paragraph{Internal Validity}
The identification of tools via a multivocal literature review entails the authors' subjective judgment when deciding whether to include sources or not and when examining the selected sources to identify the tools.
To address this possible bias, the methodology adheres to established standards in performing literature reviews, e.g., by including multiple authors in the decision process.
Still, we might not be successful in identifying all tools in our target scope, for example, because the designed methodology might not cover all relevant sources of information. 
However, the search is done on a broad body of sources and is extensive enough to reasonably claim to provide a comprehensive overview of all relevant sources.
The measurement of the tools' effectiveness will rely on manual work, both in the creation of the ground truth and in the analysis of the outputs.
As for the identification of tools, this carries the risk of biases, which we try to mitigate by including multiple authors.
At this point of the study, not all details of this step can be predicted.
Consequently, this threat to validity will be discussed further after conducting the study.

\vspace{-1mm}
\paragraph{External validity} 
We will compare the identified tools based on their measured effectiveness in architecture extraction.
This measurement could show different results in other setups.
The used dataset contains mainly models of showcase applications of small to medium size, which are rather homogeneous concerning their architectures and the used technologies.
Thus, the identified tools could perform differently on another dataset and other conclusions could be drawn concerning their comparison.
However, the dataset is the largest one currently available in the literature for this purpose.
Future work could include the creation of a dataset containing more industry-near applications, or a replication of the work if such a dataset is published by others.
The human factor in executing the tools could also affect the external validity.
We plan to mitigate this factor with the designed methodology, where extensive and equal effort will be put into attempting to run each tool.

\section{Conclusion}
This paper presents a study that we plan to conduct to identify and compare static analysis architecture recovery tools for microservice applications.
To identify existing tools, we will replicate an existing study~\cite{Bakhtin23_tools_study}, re-run it to identify new tools that appeared since its publication, and extend it into a multi-vocal literature review.
The comparison will be based on executing all identified tools on a common dataset.
Such an overview of tools and their effectiveness in architecture recovery has not been presented before, to the best of our knowledge.
The results can have implications for researchers and practitioners alike by presenting and making accessible the state-of-the-art and its capabilities, as well as by identifying gaps and thereby future research directions.

\section*{Acknowledgements}
This work is based on work supported by a grant from the Research Council of Finland (grants n. 349487 and 349488 - MuFAno) and by the National Science Foundation under Grant No. 2245287.

\bibliographystyle{ACM-Reference-Format}
\bibliography{tool_comparison}

\end{document}